\documentclass[vecphys]{svmult}
\usepackage{makeidx}         % allows index generation
\usepackage{graphicx}        % standard LaTeX graphics tool
\usepackage{natbib}        % standard LaTeX graphics tool
\usepackage{multicol}        % used for the two-column index
\usepackage[bottom]{footmisc}% places footnotes at page bottom

\newcommand\aj{{AJ}}%          % Astronomical Journal
\newcommand\araa{{ARA\&A}}%    % Annual Review of Astron and Astrophys
\newcommand\apj{{ApJ}}%        % Astrophysical Journal
\newcommand\apjl{{ApJ}}%       % Astrophysical Journal, Letters
\newcommand\apjs{{ApJS}}%      % Astrophysical Journal, Supplement
\newcommand\aap{{A\&A}}%       % Astronomy and Astrophysics
\newcommand\mnras{{MNRAS}}%    % Monthly Notices of the RAS
\newcommand\pasp{{PASP}}%      % Publications of the ASP
\newcommand{\samename}{\vrule height0.4pt depth0.0pt width1.0in \thinspace.}
\makeindex             % used for the subject index
\begin{document}

\title*{UV excess and AGB evolution in elliptical-galaxy stellar populations}
\titlerunning{UV excess in elliptical galaxies}
\authorrunning{R. A.\ Gonz\'alez-L\'opezlira \& A. Buzzoni}
\author{Rosa A.\ Gonz\'alez-L\'opezlira\inst{1}\and Alberto Buzzoni\inst{2}}
\institute{Centro de Radioastronom\'{\i}a y Astrof\'{\i}sica, Universidad 
Nacional Aut\'onoma de M\'exico, 58190 Morelia, Michoac\'an, M\'exico
\texttt{r.gonzalez@astrosmo.unam.mx}
\and INAF -- Osservatorio Astronomico di Bologna, Via Ranzani 1, 40127 Bologna, Italy \texttt{alberto.buzzoni@oabo.inaf.it}}

\maketitle

\begin{abstract}
The puzzling origin of the ``UV-upturn'' phenomenon, observed in some elliptical galaxies,
has recently been settled by identifying hot HB stars as main contributors to galaxy ultraviolet 
luminosity excess. While a blue HB morphology seems a natural 
characteristic of metal-poor stellar populations, its appearence in metal-rich 
systems, often coupled with a poorer rate of planetary nebulae per unit galaxy luminosity,
might be calling for an intimate connection between UV excess and AGB properties in early-type galaxies. 
In this work, we want to briefly assess this issue relying on infrared surface 
brightness fluctuactions as a powerful tool to trace AGB properties in 
external galaxies with unresolved stellar populations.
\end{abstract}

\section{Introduction}
\label{intro}

The so-called ``UV-upturn'' phenomenon \citep{code79}, i.e., the rising
ultraviolet emission shortward of 2000 \AA, sometimes seen in the
spectral energy distribution of elliptical galaxies and the bulges of
spirals, has been for long a puzzling problem for old galaxy environments 
dominated by stars of mass comparable to that of the Sun.

Spectroscopy and imaging \citep{brown97,brown00} 
of resolved color--magnitude (c-m) diagrams of stellar populations in
M32 have definitely shown that this UV excess traces the presence of long-lived 
O-B stars, hotter than 30\,000 - 40\,000~K  ---mostly 
hot HB stars further complemented, to a
lesser extent, by a post-AGB contribution of PN nuclei. 
However, at least two important issues need to be properly addressed to understand 
the real nature of the UV-upturn phenomenon.

{\it (i)} For such hot HB stars (naturally residing in old, metal-poor
globular clusters) to be an output of super metal-rich environments, theory requires
a quite delicate ``fine tuning'' of metallicity and stellar mass-loss 
efficiency, in order to achieve {\it (a)} the suitable partition between stellar internal He core 
and external envelope, and {\it (b)} a total mass 
$M_*$$_<\atop^{\sim}$0.52~$M_\odot$ at the HB onset \citep{dorman,castellani92}.

{\it (ii)} Observations indicate that metal-rich ellipticals display, at the same 
time, a stronger UV-upturn \citep{burstein88} {\it and} a poorer planetary nebula (PN) 
production per unit galaxy luminosity \citep{buzzoni06}. If the PN event is the final fate 
of AGB stars at the end of their thermal pulsing phase \citep{ir83}, then a PN deficiency 
might be evidence of an incomplete (or fully inhibited) AGB evolution of low-mass 
stars under special environment conditions of the parent galaxy.

As a central issue in this discussion, {\it it is clear, therefore, that a 
preeminent connection should exist between UV excess and AGB distinctive 
properties of stellar populations in early-type galaxies.}

\section{Infrared surface-brightness fluctuations as AGB probes}
\label{fluc}

\citet{tsch} first realized the potentially useful information, about their 
stellar populations, hidden in the surface brightness fluctuations (SBFs) of 
external galaxies with unresolved stellar populations. This problem has since received a more general theoretical 
assessment \citep{buzzoni93,cervino02}, exploring in many of its different facets the basic 
relationship of the theory:
\begin{equation}
{{\sigma^2(L_{\rm gal})}\over {L_{\rm gal}}} = {{\sum \ell_*^2}\over {\sum \ell_*}} = \ell_{\rm eff}.
\label{eq:fluc}
\end{equation}
The l.h. side of the equation links an observable quantity (namely, the relative variance of the
galaxy surface brightness) with the second-order statistical moment of the composing stars.
The r.h. side of eq.~(\ref{eq:fluc}) is a natural output of any population synthesis code,
and can easily be computed for different photometric bands and distinctive evolutionary
properties of simple stellar populations (SSPs).

Because of the quadratic $\ell_*$ dependence of the numerator summation in eq.~(\ref{eq:fluc}),
the effective luminosity (and its derived effective magnitude, 
$\overline{M} = -2.5\log \ell_{\rm eff}+ {\rm const}$) at any optical or near-infrared band is highly 
sensitive to the giant stars, and only marginally responds to a change in the IMF slope 
\citep[see, e.g.,][]{buzzoni93}.
Likewise, as probe of the brightest stars in a population at a given wavelength,
the effective magnitude is relatively insensitive to underlying older populations
in the case of composite systems.\footnote{For an observational confirmation, see 
\citet{gonz05a}.}

\begin{figure}[t]
\centering
\includegraphics[height=0.5\hsize]{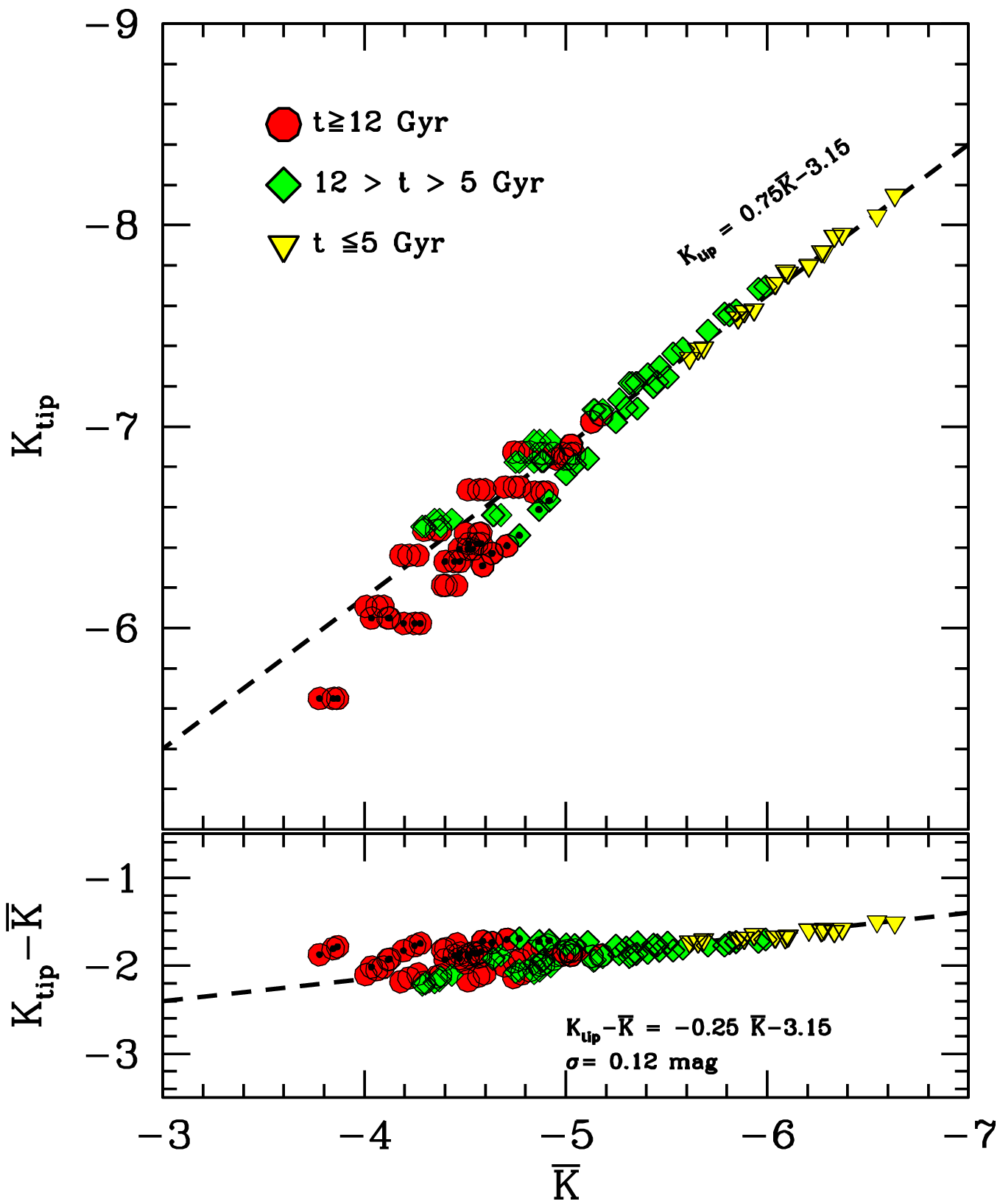}
$\qquad$
\includegraphics[height=0.52\hsize,width=0.41\hsize]{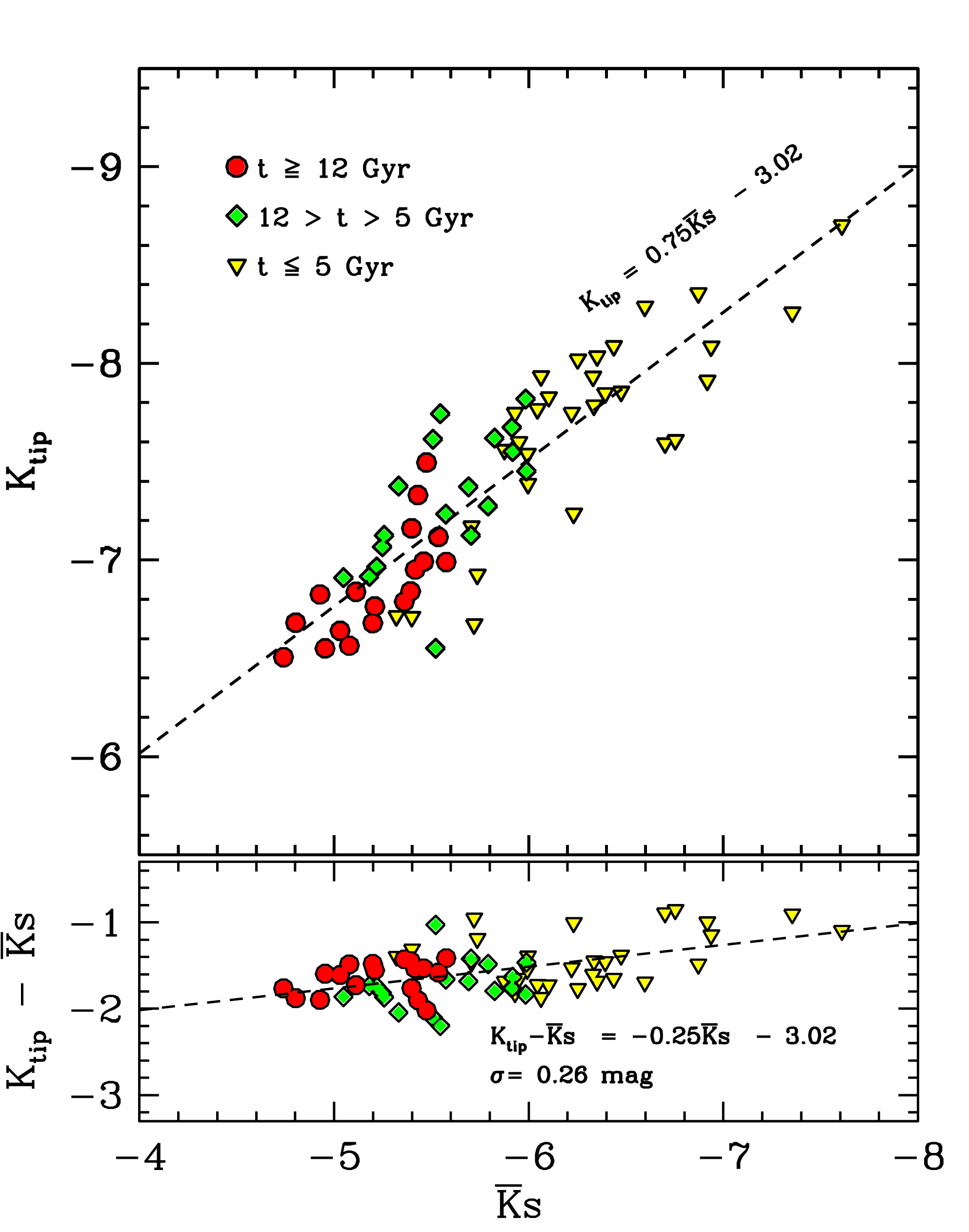}
%\vspace*{-1.6cm}
\caption{
Theoretical relationship between infrared effective magnitude ($\overline{K}$) and 
AGB tip luminosity of composing stars ($K_{\rm tip}$), for a full collection 
of SSP models from the \citet[][{\it left panel}]{buzzoni89} and Charlot \& Bruzual (2007, 
in preparation, {\it right panel}). SSP ages span from 1 Gyr (top right) to 
18 Gyr (bottom left), with a metallicity range $ 0.0001 \leq Z \leq 0.05$. The \citet{buzzoni89} models
are for a fixed \citet{reimers} mass-loss parameter $\eta = 0.3$ and a Salpeter IMF, while
the Charlot \& Bruzual (2007) calculations use a \citet{chab03} IMF, 
and inherit from \citet{mari07} the formalisms for 
$\dot M$ derived from pulsating dust-driven wind models of AGB stars.  
The tight correlation between $\overline{K}$ and $K_{\rm tip}$ appears quite 
robust along the entire range of explored parameters and, remarkably enough, nearly model 
independent.}
\label{fig:1}  
\end{figure}

As statistically representative of the stellar system as a whole, $\overline{M}$ cannot
be physically associated to any specific star or stellar group along the c-m diagram of a stellar
aggregate; nonetheless, it can be instructive to assume $\overline{M}$ as the  
magnitude of the ``prevailing'' stars in the population at the different photometric bands. 
In this respect, $\overline{K}$ 
is potentially the best tracer of the SSP tip stellar luminosity ($K_{\rm tip}$),
since {\it both quantities are expected to depend in quite the same way on the overall
distinctive parameters of the stellar population, including age, metallicity, IMF, and mass-loss}
(see Fig.~1).

The tight relationship predicted by population synthesis theory is nicely confirmed when comparing 
with real stellar clusters spanning a wide range of evolutionary parameters, like for instance 
in the Magellanic Clouds (MC). Actually, age of star clusters around these galaxies is distributed over four orders 
of magnitude, with objects as young as a few Myr, and as old as $\sim 10^{10}$~yr.
In addition, for these resolved stellar systems we can easily and consistently determine both
the fluctuation luminosity {\em and} the tip of the red giant stars from direct inspection of their
c-m diagrams.
The results of such an experiment are shown in the left panel of Fig.~2, for the sample of 191 MC star clusters 
of \citet{gonz04,gonz05}. In fact, note from the figure that the claimed relationship 
between $K_{\rm tip}$ and effective $\overline K$ magnitude is in place along the entire age range, 
and even for ultra-young (classes pre-SWB and SWB I) MC clusters, that would barely sport any standard AGB or RGB 
phases...! 

On the other hand, it is definitely worth stressing that the link between $\overline{K}$ and
$K_{\rm tip}$ is  {\it a much more deeply intrinsic property of SSPs}, and not exclusively 
related to age evolution. For instance, if one explores the impact of stellar mass-loss via
a \citet{reimers} standard parameterization (right panel of Fig.~2), we have that
quite the same relationship is still found.

\begin{figure}[t]
\centering
\centering
\includegraphics[height=0.50\hsize]{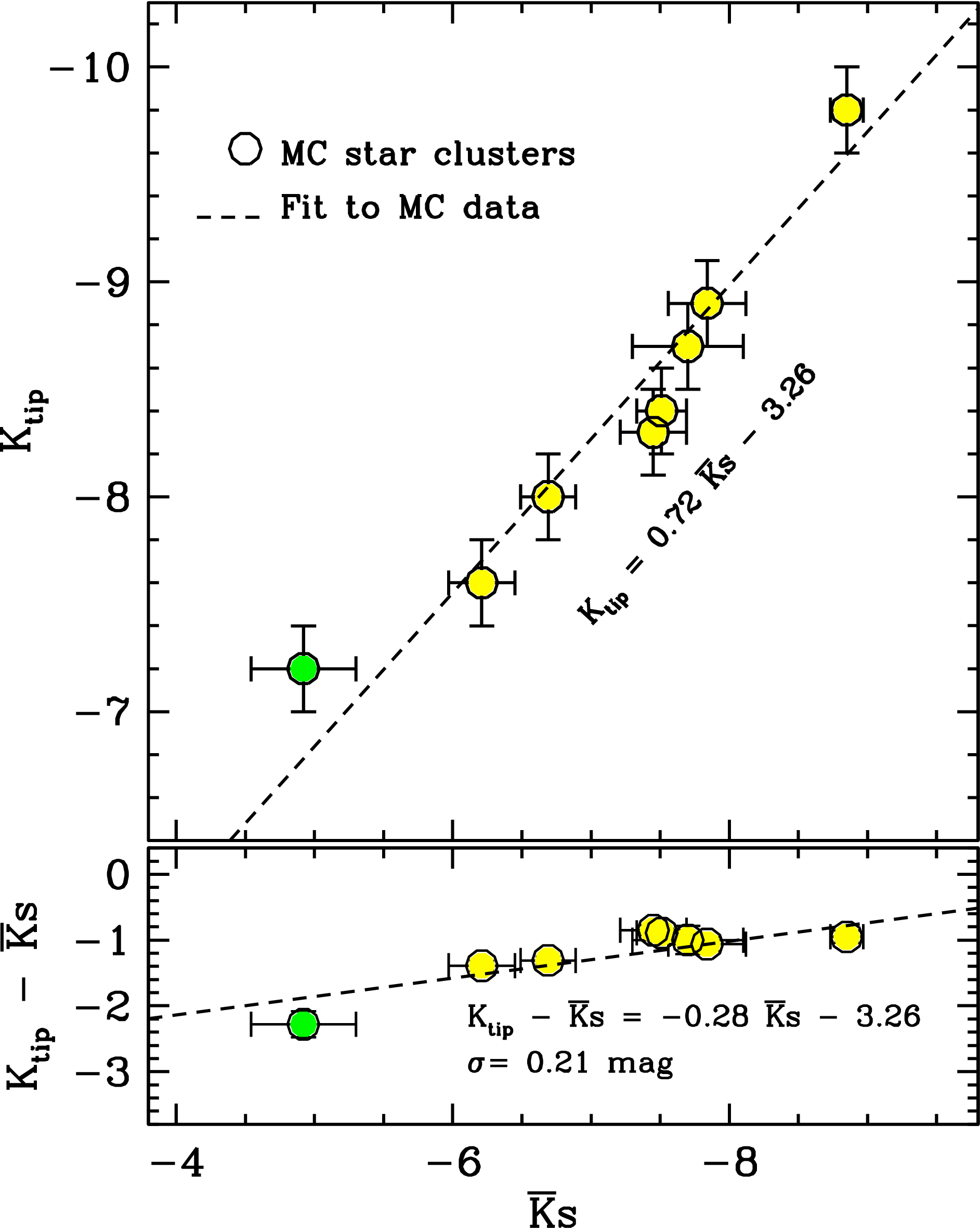}
$\qquad$
\includegraphics[height=0.51\hsize]{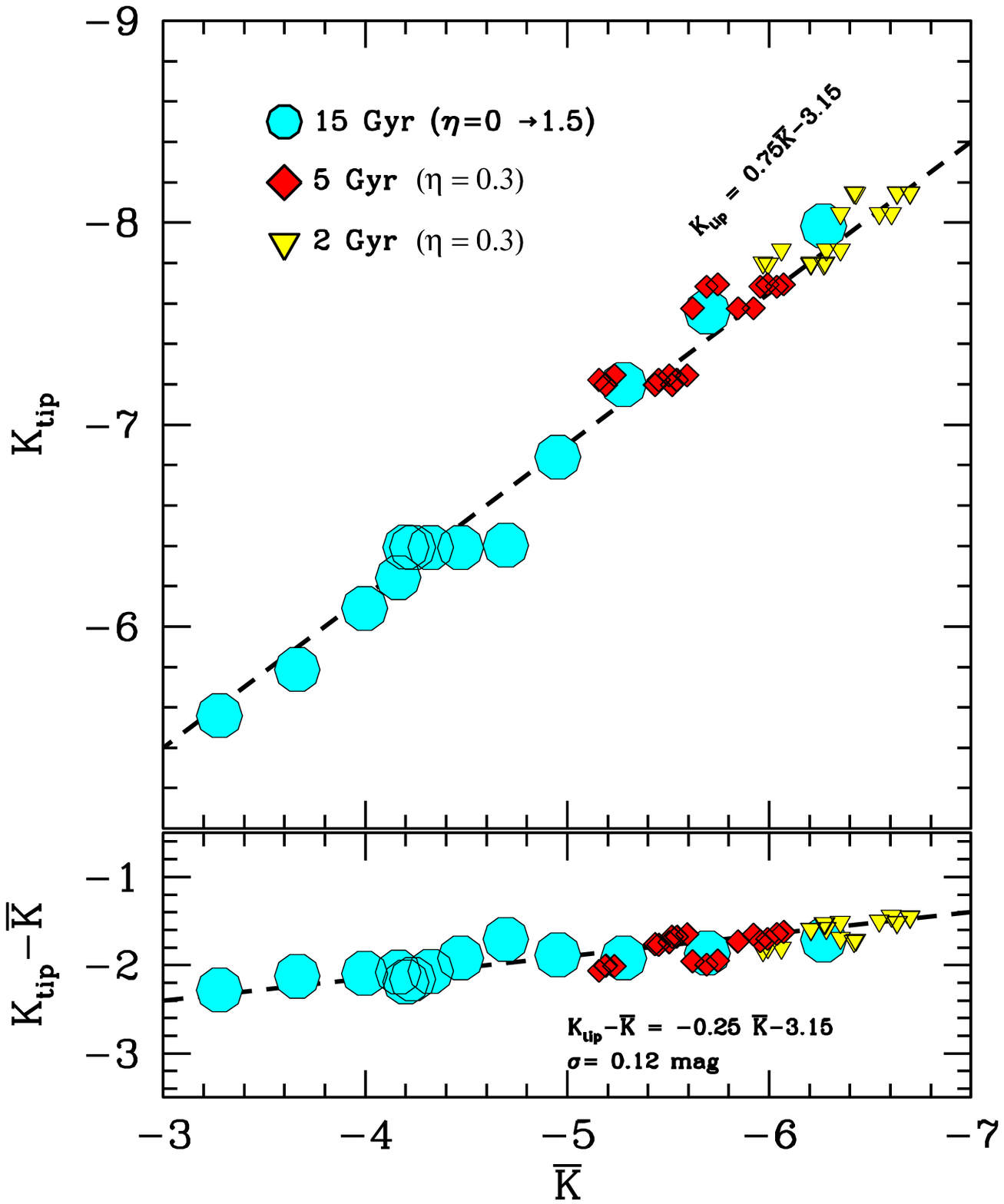}
\caption{
{\it Left panel:} $K_{\rm tip}$ magnitude vs.\ $K$-band SBFs for a sample of 191 MC star clusters from \citet{gonz04,gonz05}. 
To reduce stochastic effects due to small number stellar statistics in fast evolutionary
phases, like the AGB, the eight ``superclusters'' shown in the figure have been set up 
by coadding objects within homogeneous age bins (color-coded as in Fig.\ 1), according to classes I -- 
VII in the \citet[][SWB]{sear80} classification scheme (with older clusters bottom left in the plot), 
plus an ultra-young (pre-SWB class) supercluster (second from top right data point). 
The stellar tip luminosity has been computed from the coadded luminosity function 
of each supercluster, following \citet{lee93}. 
{\it Right panel:} big dots trace the change in a 15 Gyr SSP of solar metallicity, 
from \citet{buzzoni89}, when \citet{reimers} mass-loss parameter $\eta$ increases from 0 (top right) 
to 1.5 (bottom left). Note that a vanishing mass-loss (that is, $\eta \rightarrow 0$) 
drives the old SSP to closely resemble much younger ``standard'' cases (i.e., 2-5 Gyr SSPs 
with $\eta = 0.3$).}
\label{fig:2}       % Give a unique label
\end{figure}

\section{UV upturn and AGB extension in elliptical galaxies}\label{agbuv}

A suitable way to size up the strength of the UV-upturn in elliptical galaxies is via 
the integratad $(1550-V)$ color \citep{burstein88}, that is basically a measure of the relative 
galaxy luminosity at 1550~\AA\ compared to the visual band.\footnote{So, UV-upturn ellipticals 
have ``bluer'' [i.e., $(1550-V) \to 0$] color.} 
Figure~3 is a striking summary of the situation, as far as the comparison 
between UV-upturn and  
infrared SBF for a number of elliptical galaxies in Virgo, Leo, and
the Local Group.
A nice sequence is evident between UV excess and $K$-band effective
magnitude; in particular, UV-enhanced galaxies are about 1 mag fainter 
in $\overline K_s$, compared to UV-poor ``standard" systems.
Once translated into tip stellar luminosity (via Fig.~1), the  brightest red giants 
in UV-upturn ellipticals turn out to be about $K_{\rm tip} \simeq -7.0$, or roughly 0.7 mag fainter 
than those in UV-poor systems. 

\begin{figure}[t]
\begin{minipage}{0.68\hsize}
\includegraphics[width=\hsize]{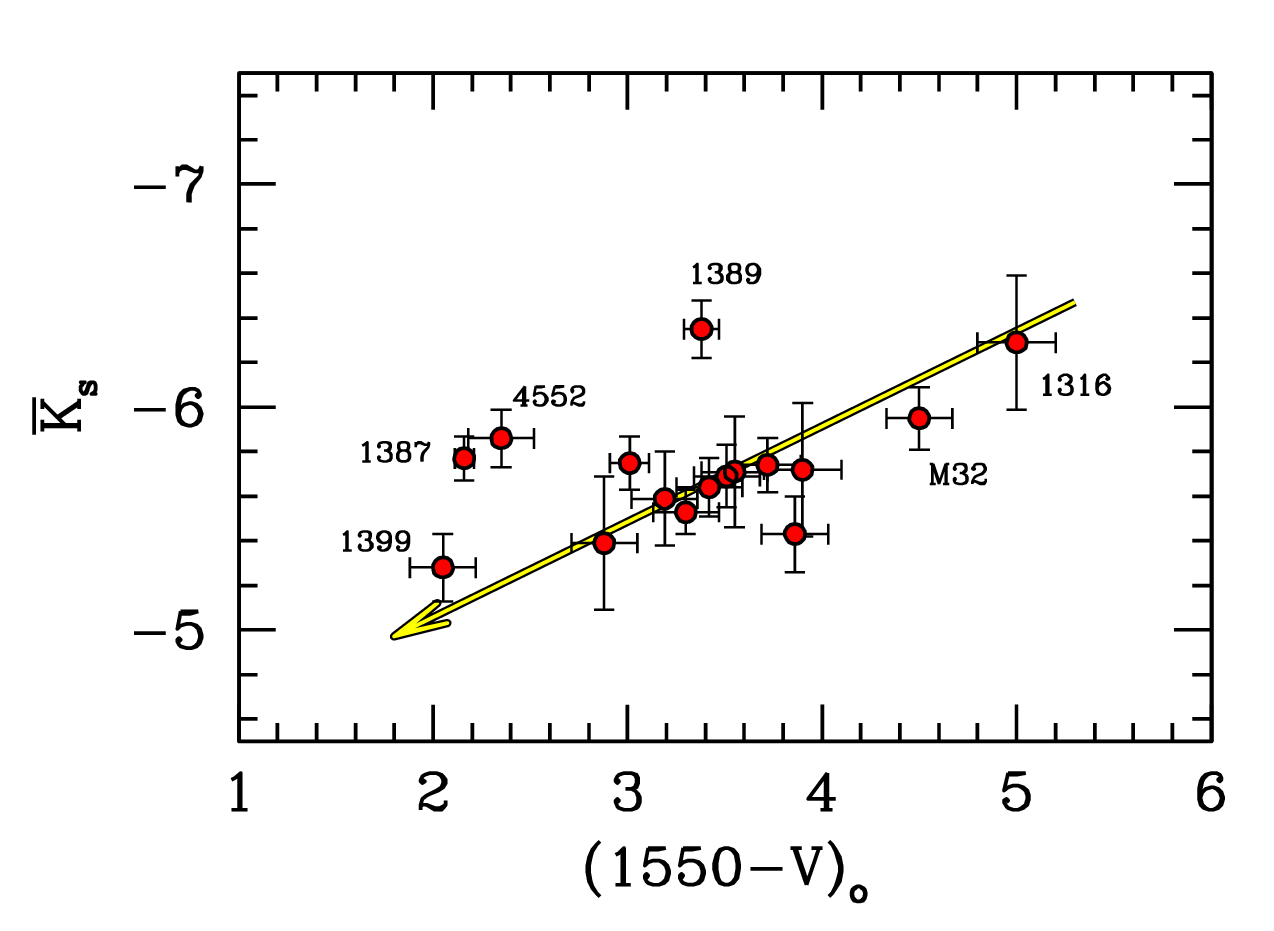}
\end{minipage}
\begin{minipage}{0.3\hsize}
\caption{
The $K$-band SBF magnitudes vs.\ (reddening cor\-rec\-ted) ultraviolet color $(1550 - V)$, for a 
sample of elliptical galaxies in the Virgo cluster, the Leo group, and the Local Group (from Buzzoni \& 
Gonz{\'a}lez 2007, in preparation).}
\end{minipage}
\vspace*{-0.6cm}
\label{fig:3}       % Give a unique label
\end{figure}

\begin{figure}[b]
\begin{minipage}{0.67\hsize}
\includegraphics[width=\hsize]{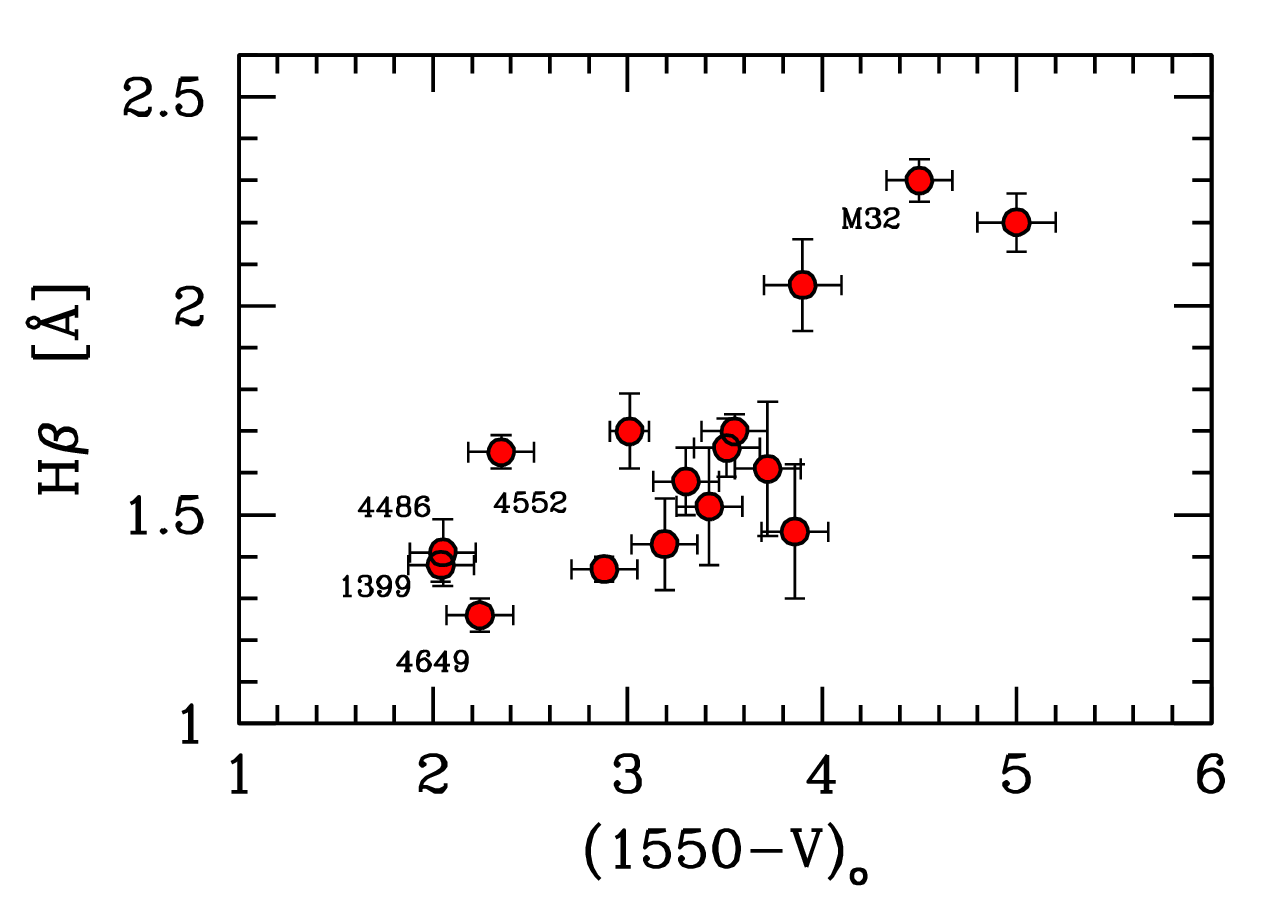}
\end{minipage}
\begin{minipage}{0.32\hsize}
\caption{
Lick $H\beta$ index vs.\ $(1550 - V)$ color for the elliptical galaxies of
Fig.~3. As $H\beta$ basically probes the temperature location of the turn-off point 
of galaxy main sequence stars \citep{buzzoni94,jens03}, the inferred physical trend
confirms that UV-upturn is stronger among $H\beta$-poor old and super metal-rich
galaxies.}
\end{minipage}
\label{fig:4}	    % Give a unique label
\end{figure}

At this stage, a number of interesting conclusions can be drawn from the analysis 
of the figure and the overall interpretative framework depicted by the synthesis model scenario.

{\it (i)} If we entirely ascribe the dimming of the red giant tip luminosity of Fig.~3 to mass-loss, 
then there should be a spread $\Delta \eta \simeq 0.4$ among the early-type galaxy population, with 
UV-enhanced ellipticals consistent with a Reimers parameter $\eta = 0.3$-0.4, as found for Galactic
globular clusters \citep{rffp}. However, this is likely an upper limit to the 
mass-loss rate range. Although elliptical's high metallicity may be 
correlated with a higher mass-loss rate at a fixed age \citep[e.g.,][]{groe95}, 
$\dot M$ certainly diminishes with older age; 
and metal-rich giant ellipticals are expected to be, on average, older than low-mass metal-poor 
systems \citep[e.g.][]{pahre98,gonz05a}, as predicted by a standard monolithic scenario \citep[see, e.g.,][]{lars74} and
directly inferred, for example, by the observed correlation between galaxy $H\beta$ index and 
$(1550-V)$ color, like in Fig.~4.

{\it (ii)} The involved $K$-band tip luminosity definitely implies that, in UV-upturn galaxies, 
the AGB luminosity extension barely exceeds the RGB tip, and only for UV-poor systems we have
to expect the branch to fully deploy, with stars 1-1.5~mag brighter than the RGB tip 
\citep[][]{buzzoni95}.

{\it (iii)} A more careful comparison with the evolutionary synthesis models indicates 
that, on average, AGB stars in ellipticals always experience the thermal-pulsing
phase likely giving rise to the PN event. However, given a reduced AGB extension
among UV-enhanced galaxies, this situation is increasingly countered when either galaxy age or 
metallicity increase, and a higher number of stars end up as UV-bright {\it AGB-manqu\'e} objects 
\citep[][]{greggio}, thus escaping the PN phase and directly feeding the high-temperature 
region of the galaxy c-m diagram.

\vspace*{0.3cm}
We would like to thank Gustavo Bruzual for providing us with the latest Charlot \& Bruzual (2007) 
SSP models, in advance of publication. 

%
%%%%%%%%%%%%%%%%%%%%%%%%%%%%%%%%%%%%%%%%%%%%%%%%%%%%%%%%%%%%%%%%%%%

%%%%%%%%%%%%%%%%%%%%%%%%%%%%%%%%%%%%%%%%%%%%%%%%%%%%%%%%%%%%%%%%%%%%%%  }

\printindex
\end{document}